\documentclass[preprint,aps,showpacs,nofootinbib]{revtex4}
\usepackage{epsfig,amssymb,amsmath}

\def\c{\varepsilon}

\def\v{\varphi}

\newcommand{\beq}{\begin{equation}}
\newcommand{\eeq}{\end{equation}}
\newcommand{\bea}{\begin{eqnarray}}
\newcommand{\eea}{\end{eqnarray}}
\newcommand{\bear}{\begin{array}}
\newcommand {\eear}{\end{array}}
\newcommand{\bef}{\begin{figure}}
\newcommand {\eef}{\end{figure}}
\newcommand{\bec}{\begin{center}}
\newcommand {\eec}{\end{center}}
\newcommand{\non}{\nonumber}

\newcommand{\la}{\left\langle}
\newcommand{\ra}{\right\rangle}
\newcommand{\ds}{\displaystyle}

\def\GEV#1{10^{#1}{\rm\,GeV}}

\def\lrfp#1#2#3{ \left(\frac{#1}{#2} \right)^{#3}}
\def\oten#1{ {\mathcal O}(10^{#1})}

\begin{document}
\draft
\tighten
\preprint{TU-943,~~
IPMU13-0159}
\title{\large \bf
New inflation in supergravity after Planck and LHC
}
\author{
      Fuminobu Takahashi$^{(a,b)}$\footnote{email: fumi@tuhep.phys.tohoku.ac.jp}}
\affiliation{
$^{(a)}$Department of Physics, Tohoku University, Sendai 980-8578, Japan,\\
$^{(b)}$Kavli Institute for the Physics and Mathematics of the Universe (WPI), TODIAS, University of Tokyo, Kashiwa 277-8583, Japan
    }

\vspace{2cm}

\begin{abstract}
We revisit a single-field new inflation model based on a discrete $R$ symmetry. Interestingly,
the inflaton dynamics naturally leads to a heavy gravitino of  mass $m_{3/2} = {\cal O}(1-100)$\,TeV,
which is consistent with the standard-model like Higgs boson of mass $m_h \simeq 126$\,GeV.
However, the predicted spectral index $n_s \approx 0.94$ is
 in tension with the Planck result, $n_s = 0.9603\pm0.073$. We show that
the spectral index can be increased by allowing a small constant term in the superpotential during inflation.  
The required size of the constant is close to the largest allowed value for successful inflation, and it may
be a result of a pressure toward larger values in the landscape. Alternatively,
such constant term may arise in association with supersymmetry breaking  required to
cancel the negative cosmological constant from the inflaton sector. 
\end{abstract}

\maketitle


\section{Introduction}
\label{sec:1}
The observed temperature correlation of the cosmic microwave background radiation (CMB) strongly 
suggests that our Universe experienced an accelerated expansion, i.e., inflation~\cite{Guth:1980zm,Linde:1981mu},  
at an early stage of the evolution. Recently, the Planck
satellite measured the CMB temperature anisotropy with an unprecedented accuracy, and 
tightly constrained properties of the primordial density perturbation such as the spectral index ($n_s$),
the tensor-to-scalar ratio ($r$), non-Gaussianities, etc.~\cite{Ade:2013zuv}. 

The tensor-to-scalar ratio, if discovered, would pin down the inflation energy scale. 
Up to now, there are various models of chaotic inflation developed in 
supergravity~\cite{Kawasaki:2000yn,Kawasaki:2000ws, Kallosh:2010ug,Kallosh:2010xz,Takahashi:2010ky,Nakayama:2010kt,
Harigaya:2012pg,Nakayama:2013jka,Croon:2013ana} and string theory~\cite{Silverstein:2008sg,McAllister:2008hb}.
On the other hand, if the tensor mode is not detected in the future observations, 
it will give preference to low-scale inflation models.  Then it is the spectral index that can be used to select inflation models.
The Planck result for the spectral index is given by~\cite{Ade:2013zuv}
\beq
n_s \;=\; 0.9603 \pm 0.0073 ~~(68 \%\,{\rm errors};\,{\rm Planck + WP}),
\label{ns}
\eeq
which already constrains various low-scale inflation models such as a hybrid inflation model~\cite{BasteroGil:2006cm}.

One of the low-scale inflation models is the so called new inflation~\cite{Linde:1981mu}. We consider a single-field new inflation model
based on a discrete $R$ symmetry in supergravity, proposed in Refs.~\cite{Kumekawa:1994gx,Izawa:1996dv}.
This inflation model is simple and therefore attractive; it contains only a single inflaton field, whose interactions are 
restricted by the discrete $R$ symmetry.\footnote{ The same model was considered in Ref.~\cite{Dine:2011ws}.
See e.g. Refs.~\cite{Dine:2009swa,Dine:2010eb,Harigaya:2013vja}  for other phenomenological and cosmological study of the discrete R symmetry.
} Intriguingly, the inflation dynamics necessarily  leads to a non-zero 
vacuum expectation value of the superpotential after inflation, giving the dominant contribution to the gravitino mass in the low energy. 
As we shall see shortly, it can naturally explain the gravitino mass of ${\cal O}(1 - 100)$\,TeV, which is consistent
with the observed standard-model like Higgs boson of mass $m_h \simeq 126$\,GeV~\cite{Giudice:2011cg}. 
However, it is known that the predicted spectral index is approximately given by $n_s \approx 0.94$ 
for the e-folding number $N_e = 50$, which has a  tension with the Planck result (\ref{ns}).

In this letter we will show that the spectral index in the single-field new inflation model 
is increased if one allows a small but non-zero
constant term in the superpotential during inflation. It was known that  the constant term during inflation cannot 
give the dominant contribution to the gravitino mass in the low energy, since otherwise it would spoil the inflaton dynamics~\cite{py}. 
However, its effects on the inflaton dynamics,
especially on the spectral index, has not been studied so far.\footnote{
The effect of such constant superpotential on the inflation dynamics in  hybrid inflation was considered
in Refs.~\cite{Nakayama:2010xf,Buchmuller:2000zm}.
} 
We will show that the $n_s$ in the range of (\ref{ns}) 
can be realized, and derive an upper bound on $n_s$ that can be reached by
adding such constant term.

\section{Single-field new inflation model in supergravity}
\label{sec:2}

Let us consider a new inflation model given in Refs.~\cite{Kumekawa:1994gx,Izawa:1996dv,Ibe:2006fs} as an
example of  low-scale inflation models. The K\"ahler potential and
superpotential of the inflaton sector are written as
\bea
K(\phi,\phi^\dag) &=& |\phi|^2 + \frac{k}{4 } |\phi|^4 + \cdots,\non\\
W(\phi) &=&  v^2\phi -  \frac{g}{n+1}\, \phi^{n+1} + c,
\label{IY}
\eea
where $\phi$ is the inflaton superfield, the dots represent higher-order terms that are irrelevant for our purpose, and 
$k$ and $g$ are coupling constants of order unity. Here and in what follows we
adopt the Planck units in which the reduced Planck mass $M_p \simeq 2.4 \times \GEV{18}$
is set to be unity. The structure of the superpotential can be understood in terms of a discrete
R-symmetry $Z_{2n}$ under which $\phi$ has a charge $2$.

The main difference of (\ref{IY}) from the original model in 
Refs.~\cite{Kumekawa:1994gx,Izawa:1996dv} is that we have included
a constant term in the superpotential, $c$, which breaks the discrete R symmetry.
It can be originated from various sources such as gaugino condensation, supersymmetry (SUSY) breaking, 
and flux compactifications~\cite{Gukov:1999ya}, and therefore, it is no wonder if such constant
term exists during inflation. 
It is known that $c$ cannot give the 
dominant contribution to the gravitino mass in the low energy, since otherwise it would spoil
the inflation~\cite{py}. The purpose of this letter is to study its effect on the inflaton dynamics when $c$ is
sufficiently small. For simplicity we take all the parameters real and positive and set the
e-folding number $N_e = 50$. This greatly simplifies the analysis of the inflaton dynamics
as it is effectively described by a single-field inflation model. A full analysis for the case of a complex
$c$ will be presented elsewhere. 

The inflaton potential in supergravity is given by
\beq
V(\phi) = e^K \left(D_\phi W K^{\phi {\bar \phi}} (D_\phi W)^* - 3 |W|^2 \right)
\eeq
with $D_\phi W = W_\phi + K_\phi W$.  In terms of the real component, $\v = \sqrt{2}{\rm\,Re} [\phi]$,
the inflaton effective potential can be approximately expressed as
\beq
V(\v) \;\simeq\; v^4 -2 \sqrt{2}\, c v^2 \v- \frac{k}{2} v^4 \v^2 - \frac{g}{2^{\frac{n}{2}-1}} v^2 \v^n + \frac{g^2}{2^n} \v^{2n}.
\label{Vinf}
\eeq
The inflaton potential is so flat near the origin that inflation takes place, and 
the inflaton $\v$ is stabilized at 
\beq
\v_{\rm min} \;\simeq \sqrt{2} \lrfp{v^2}{g}{\frac{1}{n}}
\eeq
with the mass
\beq
m_\varphi \;\simeq\; nv^2 \lrfp{v^2}{g}{-\frac{1}{n}} \simeq (n+1)\, m_{3/2} \lrfp{\varphi_{\rm min}}{\sqrt{2}}{-2}.
\eeq
As a result,  the gravitino mass after inflation is given by
\beq
m_{3/2} = \la e^{K/2} W \ra \simeq \frac{n v^2}{n+1}  \lrfp{v^2}{g}{\frac{1}{n}}.
\eeq
This is one of the interesting features of the new inflation model; the SUSY breaking
scale is directly related to the inflation dynamics.\footnote{We assume that  SUSY breaking
in a hidden sector provides a positive contribution to the cosmological constant, so that the total cosmological
constant almost vanishes. See also discussion in Sec.~\ref{sec:4}.
} In particular, the gravitino mass falls in the range
of ${\cal O}(1 - 100)$\,TeV for $n=4$, which is consistent with the observed standard-model like 
Higgs boson of mass $m_h \simeq 126$\,GeV~\cite{Giudice:2011cg}. The cosmological and phenomenological
implications of the SUSY breaking scale of this order were studied extensively in the literature~\cite{Ibe:2011aa}.
We assume that the constant term $c$ is
so small that it does not affect the inflaton potential minimum as well as the gravitino mass
in the low energy. Nevertheless, as we will see shortly, it can have a significant effect on the
inflaton dynamics during inflation as well as the predicted spectral index. 

The presence of the linear term in the inflaton potential (\ref{Vinf}) makes it difficult to solve the
inflaton dynamics analytically, and we need to resort to a numerical treatment. The numerical results
will be presented in the next section. Before proceeding, here let us explain how
the addition of the linear term changes the predicted spectral index. At the leading order, the spectral
index is expressed in terms of the slow-roll parameters as
\beq
n_s \;\simeq\; 1  - 6 \c + 2 \eta,
\label{nsana}
\eeq
where  $\c$ and $\eta$ are defined by
\beq
\c \equiv \frac{1}{2} \lrfp{V'}{V}{2},~~~~~\eta \equiv \frac{V''}{V}.
\eeq
The prime denotes the derivative with respect to the inflaton $\varphi$, and the slow-roll parameters
are evaluated at the horizon exit of cosmological scales.
In the low-scale inflation models, the contribution of $\c$ is generically much smaller than $\eta$,
and therefore, it is the curvature of the inflaton potential that determines the spectral index. 
So, one might expect that the inclusion of the linear term hardly affects
the predicted spectral index. This is not the case, however. In fact, it significantly changes the predicted
spectral index as it affects the inflaton dynamics. 

Let us denote the inflaton field value at the horizon exit of cosmological scales
as $\varphi_N$.  In the presence of the linear term with $c >0$, the inflaton potential becomes
steeper and so $\varphi_N$ becomes smaller than without. As a result, 
the curvature of the potential becomes smaller as long as $\varphi_N > 0$, and $n_s$ becomes larger.
The largest value of $n_s \approx 1-2k$ is obtained when $\varphi_N = 0$, where the curvature of the potential comes only
from the third term in (\ref{Vinf}).   For an even larger $c$, $\varphi_N$ becomes negative, and then, the spectral index
turns to decrease. If $c$ is sufficiently large, the inflation does not last for more than $50$ e-foldings. 
We note here that the quartic coupling $k$ in the K\"ahler potential has a similar effect. As one can see
from Fig.~1 in Ref.~\cite{Ibe:2006fs}, a positive $k$ can slightly increase $n_s$. This is because $\varphi_N$
becomes smaller for larger $k$, i.e., for steeper inflaton potential. In contrast to the linear term, however,
this effect is partially canceled by its contribution to $n_s$, since $k$ directly contributes to $n_s$ through $\eta$.

How large should $c$ be in order to affect the inflaton dynamics? In the absence of the linear term,
the inflaton field value at the horizon exit, $\varphi_{N0}$,  is given by~\cite{Ibe:2006fs}
\bea
\varphi_{N0}^{n-2} &\simeq& \frac{k v^2 2^{\frac{n}{2}-1}}{ g n} 
\left(\frac{1+k(n-2)}{1-k} e^{N_e k (n-2)}-1\right)^{-1}.
\eea
In the limit of $k \rightarrow 0$, it is given by $\varphi_{N0}^{n-2} \rightarrow  \beta {\hat g}^{-1}$ 
with
\beq
\beta \;\equiv \; \frac{2^{\frac{n}{2}-1}}{g n ((n-2) N_e+n-1)} \simeq 0.049,
\eeq
where the second equality holds for $n=4$ and $N_e = 50$. For simplicity we assume $k=0$
in the following.  In order to affect the inflaton dynamics, especially when the cosmological scales exited the horizon,
the tilt of the potential should receive a sizable contribution from the linear term at $\varphi \sim \varphi_N$. Therefore,
comparing the second and fourth terms in (\ref{Vinf}), 
\beq
2 \sqrt{2} c \, v^2\gtrsim \frac{n g}{2^{\frac{n}{2}-1}} v^2 \varphi_{N0}^{n-1},
\eeq
namely,
\bea
{\hat c} &\gtrsim&
 \frac{\left((n-2)(N_e+1)+1\right)^\frac{1-n}{n-2}}{2\sqrt{2}} \lrfp{n {\hat g}}{2^{\frac{n}{2}-1}}{- \frac{1}{n-2}}
\simeq \frac{2.4 \times 10^{-4}}{\sqrt{\hat g}},
\label{hatc}
\eea
must be satisfied to affect the spectral index, where ${\hat c} \equiv c/v^2$ and ${\hat g} \equiv g/v^2$.
Here the second equality holds for $n=4$ and $N_e = 50$.
In fact, the maximum of $n_s$ is realized for ${\hat c}$ about one order of magnitude larger than the lower bound (\ref{hatc}).
Thus, in order to give sizable modifications to the spectral index, 
${\hat c}$ must be larger for a smaller ${\hat g}$, or equivalently, for a heavier gravitino mass.

\section{Numerical results}
\label{sec:3}
In this section we present our numerical results. We have solved the equation of motion for the inflaton
with the potential (\ref{Vinf}) and estimated the spectral index based on a refined version of (\ref{nsana}) including
up to the second-order slow-roll parameters. We have also imposed the Planck normalization of the
primordial density perturbations~\cite{Ade:2013zuv},
\beq
{\cal P_R} = \frac{1}{12 \pi^2} \frac{V(\varphi_N)^3}{V'(\varphi_N)^2} \simeq 2.2 \times 10^{-9}.
\eeq
at the pivot scale $k_0 = 0.05$\,Mpc$^{-1}$, by iteratively adjusting $v$ while ${\hat c} = c/v^2$ and ${\hat g} =
g/v^2$ are fixed. The reason for fixing ${\hat c}$ and ${\hat g}$ instead of $c$ and $g$ is 
to avoid the interference of the iteration procedures for the Planck normalization
in solving the inflaton dynamics. 

\begin{figure}[ht]
\begin{center}
\includegraphics[scale=1]{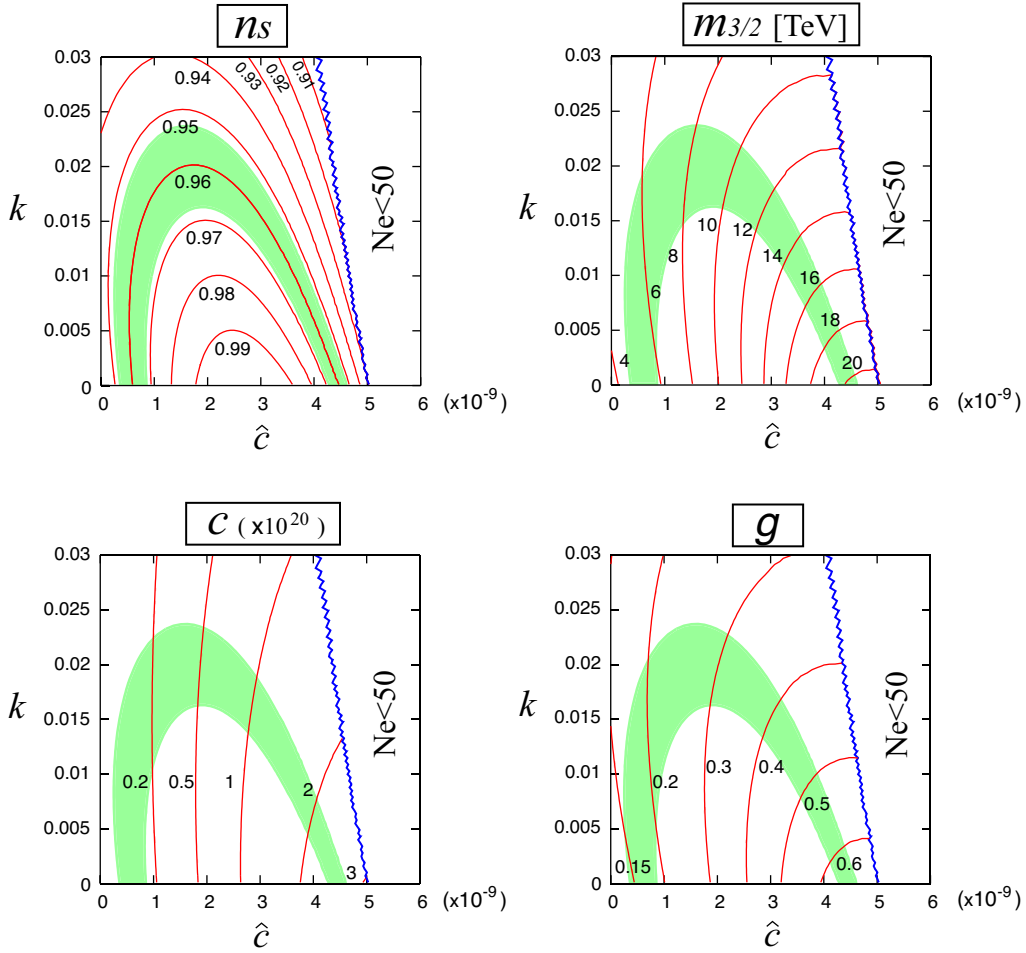}
\caption{
Contours of $n_s$, $m_{3/2}$, $c$ and $g$ in the $({\hat c}, k)$ plane
for $\varphi_{\rm min}/\sqrt{2} =3 \times \GEV{15}$. The Planck result
for $n_s$ in Eq.~(\ref{ns}) is shown in the shaded region. In the right white region,
the total e-folding number is smaller than $50$.
}
\label{fig:ns1}
\end{center}
\end{figure}

\begin{figure}[ht]
\begin{center}
\includegraphics[scale=1]{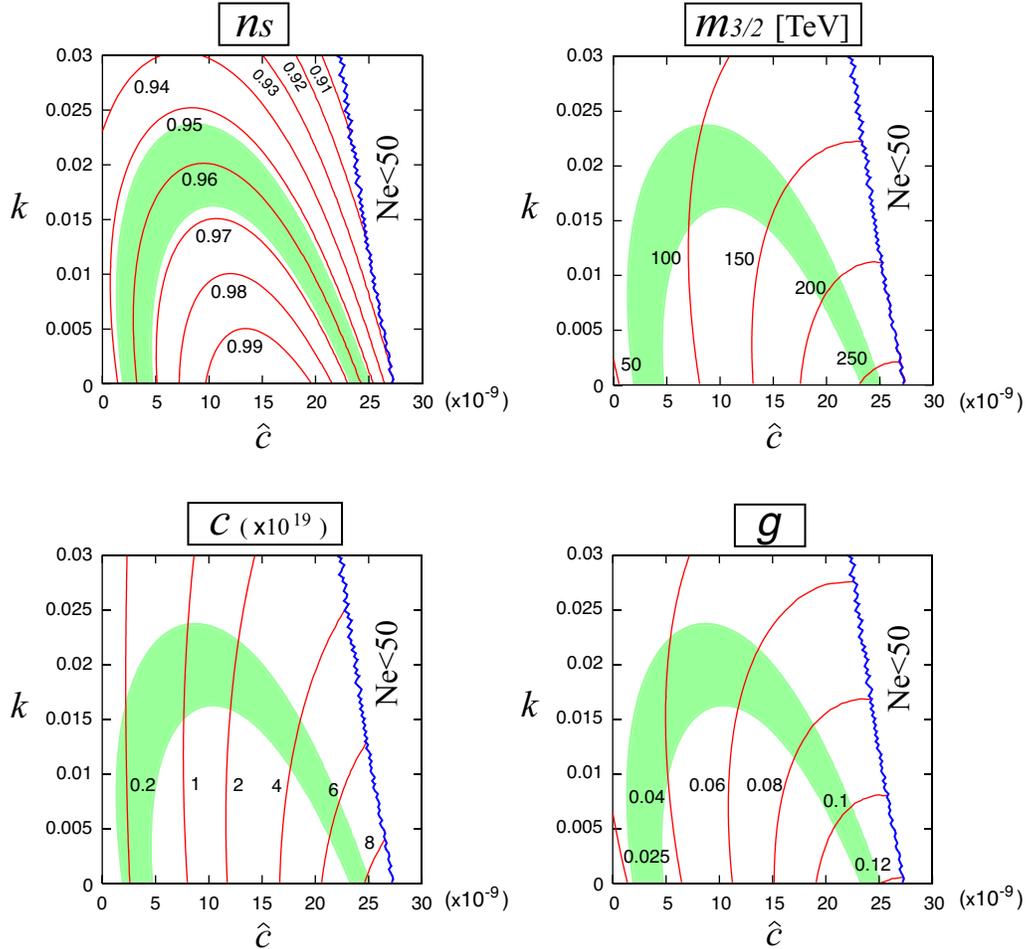}
\caption{
Same as Fig~\ref{fig:ns1} but for $\varphi_{\rm min}/\sqrt{2} = 7 \times \GEV{15}$.
}
\label{fig:ns2}
\end{center}
\end{figure}

We have shown the contours of $n_s$, $m_{3/2}$ in units of TeV, $c$ and $g$ 
in the $({\hat c}, k)$ plane for $\varphi_{\rm min}/\sqrt{2} =  3 \times \GEV{15}$ and $7 \times \GEV{15}$  
in Fig.~\ref{fig:ns1} and Fig.~\ref{fig:ns2}, respectively. (Note that the potential minimum is fixed during the
iterative procedures for the Planck normalization.)
  We can see that, for fixed $k$,  the spectral index increases as  ${\hat c}$ increases until it reaches $1-2k$, 
 and then turns to decrease since $\varphi_N$ becomes negative. The maximum of $n_s$ coincides with our
 expectation given in the previous section. 
For a sufficiently large $c$, the total e-folding number is smaller than $50$,
and the inflation model cannot account for the observed density perturbations.\footnote{
In our numerical calculation, we have required the initial position of the inflaton should be
deviated from the local maximum of the potential at least by hundred times the quantum fluctuations $\sim H_{\rm inf}/2\pi$,
for our classical treatment of the inflaton dynamics to be valid. 
}   Note also that, in contrast to the case without the linear term, the inclusion of the quartic coupling $k$ does not increase $n_s$,
as it is the linear term that affects $\varphi_N$. Most importantly, thanks to the constant term $c$,
there appeared a parameter space shown in the shaded region where the spectral index satisfies the 1 $\sigma$  region allowed by the Planck (\ref{ns}),
$n_s  = 0.9603 \pm 0.0073$.
The required value of $c$ is $c \approx 10^{-20} - 10^{-19}$, whose
implications will be discussed in the next section.

\section{Discussion and conclusions}
\label{sec:4}
We have so far assumed that $c$ is a positive and real number, which enabled us to focus on the dynamics along the real
component of the inflaton. In general, however, $c$ is complex, and  the inflaton dynamics is described by both
the real and imaginary components of $\phi$. We leave the full analysis of the general case to future work, but our 
discussion in Sec.~\ref{sec:2} should also apply to this case; the linear term can modify the inflaton dynamics such that
$|\phi|$ at the horizon exit of cosmological scales is smaller than without, and as a result the spectral index can be increased. 
We expect that there will be regions of parameter space in which $n_s$ is close to $0.96$ in the general case.

There are a variety of sources that can generate the constant term $c$ in the superpotential. 
For instance, there may be gaugino condensation in a hidden sector. The required value of $c \approx
10^{-20} - 10^{-19}$ suggests the dynamical scale $\Lambda \approx 10^{-7} - 10^{-6}$ at an intermediate scale. 

Let us now suppose that
 the dynamics leading to the constant $c$ also gives rise to the SUSY breaking F-term 
 order $\Lambda^2$:
 \beq
 W = \mu^2 Z 
 \eeq
 with $\mu^2 \sim \Lambda^2$. Here $Z$ is the SUSY breaking field.
 If $\la Z \ra \sim \Lambda$, the above superpotential leads to  $c \sim \Lambda^3$. For instance, this can be
 realized in the IYIT SUSY breaking model~\cite{IYIT}, if the origin is destabilized due to a large Yukawa coupling ($\sim 4 \pi$)
  with the dynamical quarks. To be concrete we consider the case in Fig.~\ref{fig:ns2}, in which $m_{3/2} \sim 100$\,TeV and
  $c\sim 10^{-19}$.
  In this case, it is interesting to note that  the SUSY breaking generates a positive contribution to the potential
  $\Lambda^4 \sim c^\frac{4}{3} \sim 10^{-26}$,
which can be naturally canceled by the negative contribution  from the inflaton sector $-3m_{3/2}^2 \sim -10^{-26}$. That is to say, the SUSY breaking sector  provides
both the constant term $c$ favored by observations and the required positive contribution to the cosmological constant.\footnote{We note that,
although this is an order-of-magnitude estimate, the agreement is better for $m_{3/2} \sim 100$\,TeV
compared to $m_{3/2} \sim {\cal O}(1)$\,TeV.}
Put another way, the inflaton sector provides a large constant term in the superpotential that is required
for a vanishingly small cosmological constant in the dynamical SUSY breaking scenario.\footnote{See Ref.~\cite{Dine:2006gm}
for retrofitting the O'Raifeartaigh models without such large contribution from the inflation sector. } 
Then the inflaton dynamics as well as the spectral index are necessarily 
modified from the original model in Refs.~\cite{Kumekawa:1994gx,Izawa:1996dv}.
Therefore, the predicted value of $n_s$ is naturally deviated from that in the original model without any 
fine-tuning of parameters. 

Alternatively, the size of $c$ could be determined by requiring successful inflation,
 regardless of the SUSY breaking.
Suppose that the value of $c$ is determined by some high energy physics such as flux compactifications,
and that there is a pressure toward larger values of $|c|$. As long as the new inflation is responsible
for the observed density perturbations, $|c|$ cannot be arbitrarily large as it would spoil the
inflation dynamics. Indeed, as can be seen from Fig.~\ref{fig:ns1} and \ref{fig:ns2},
the total e-folding number would be too small for a sufficiently large $|c|$. Then, under the assumption of
successful inflation, the conditioned probability distribution for $|c|$ may be biased to larger values with a sharp cut off, 
corresponding to the total e-folding number $N_e \gtrsim 50$. 
In this case the spectral index is naturally deviated from the original prediction.

The inflaton generally decays into the gravitino which potentially leads to cosmological problems.~\cite{Kawasaki:2006gs,Asaka:2006bv,Dine:2006ii,Endo:2006tf,Endo:2007ih}. Assuming that the SUSY breaking field is heavier than the inflaton, the gravitino abundance is estimated by
\beq
\frac{n_{3/2}}{s} \;\sim\;  10^{-12} \lrfp{T_R}{\GEV{7}}{-1}
\lrfp{\la \phi \ra}{7 \times \GEV{15}}{2} \lrfp{m_\varphi}{6\times \GEV{10}}{2},
\label{Y32}
\eeq
where $T_R$ is the reheating temperature of the inflaton, and $\la \phi \ra$ and
$m_\varphi$ represent the vacuum expectation value and the mass of the inflaton 
at the potential minimum. 
The values in the parenthesis are for the case of $n=4$.
The cosmological constraints on the gravitino abundance are given by
\begin{equation}
\begin{split}
 \frac{n_{3/2}}{s}\;  \lesssim \;
	\begin{cases}
	5\times 10^{-13} &~~ {\rm for}~~~10\,{\rm TeV} \lesssim m_{3/2} \lesssim 30\,{\rm TeV},\\
	\ds{4\times 10^{-13}\left( \frac{1\,{\rm TeV}}{m_{\rm LSP}} \right)}&~~ {\rm for}~~~m_{3/2} \gtrsim 30\,{\rm TeV},
	\end{cases}
	\label{gravbound}
\end{split}
\end{equation}
where $m_{\rm LSP}$ denotes the lightest SUSY particle (LSP) mass, and the R-parity conservation
is assumed.  If the R-parity is violated by a
small amount, the LSP can decay before BBN, and there will be no
cosmological constraint on the gravitino abundance for $m_{3/2} \gtrsim 30\,{\rm TeV}$.
The inflaton decay into right-handed neutrinos was
studied in Ref.~\cite{Ibe:2006fs} and the reheating temperature can be of $\oten{6-7}$\,GeV. Thus,
comparing (\ref{Y32}) with (\ref{gravbound}), one can see that the LSPs produced by gravitino decays
may be able to account for the observed dark matter abundance, if its thermal relic abundance is suppressed
as in the case of Wino LSP. This is consistent with the scenario in \cite{Ibe:2011aa}. Note that the
gravitino abundance depends on the details of the SUSY breaking sector, and it can be suppressed 
if the SUSY breaking field is much lighter than the inflaton mass~\cite{Nakayama:2012hy}.

We have focused on the case of $n=4$ in this letter. For $n \geq 5$, the predicted spectral
index is larger, and is consistent with the Planck result (\ref{ns}) even in the absence of the constant
term. The effect of the constant term will be similar in this case; if $c$ is positive (negative), it will increase
(decrease) $n_s$ as long as $\varphi_N > 0$. For $n \geq 5$, the inflaton mass, expectation value,
and the gravitino mass become larger compared to the case of $n=4$, and as a result,
more gravitinos are produced by the inflaton decay.
The LSPs  from the gravitino decay tends to be overproduced, but this problem can be avoided by 
adding a small R-parity violation.

In this letter we have revisited the single-field new inflation model in supergravity,
focusing on the case of $n=4$ in which the inflaton dynamics leads to the gravitino mass
of ${\cal O}(1-100)$\,TeV. Interestingly,
this may explain  why SUSY particles have not been observed experimentally to date. 
The predicted spectral index for this model is known to be relatively small, $n_s \approx 0.94$,
which is in tension with the Planck results. 
We have shown that the spectral index can be increased so that it lies within $1\sigma$
region allowed by the Planck data,  if there is a small but non-zero constant term
in the superpotential during inflation.  Even if it  gives only subdominant
contributions to the low-energy gravitino mass, it can have a significant impact on 
the inflaton dynamics, and therefore the spectral index. Interestingly, the required size
of the constant term is close to the largest allowed value for successful inflation.
It may be a result of the pressure toward higher values of $|c|$. Alternatively, it 
can be easily generated by hidden sector dynamics associated
with the SUSY breaking, which  cancels the negative contribution from the inflaton sector for
a vanishingly small cosmological constant.

\section*{Acknowledgment}
The author thanks Michael Dine for communications at Tohoku Workshop on ``Higgs and Beyond" where 
the present work was initiated.  The author is grateful to T.~Yanagida for useful comments.
This work was supported by
Grant-in-Aid for  Scientific Research on Innovative
Areas (No.24111702, No. 21111006, and No.23104008), Scientific Research (A)
(No. 22244030 and No.21244033), and JSPS Grant-in-Aid for Young Scientists (B)
(No. 24740135), and Inoue Foundation for Science.
This work was also supported by World Premier International Center Initiative
(WPI Program), MEXT, Japan [FT].

\end{document}